\documentclass{icrc2009}
\usepackage{url}
\usepackage{graphicx}   
\usepackage[caption=false]{caption}    
\usepackage[font=footnotesize]{subfig} 
\usepackage{fixltx2e}

\newcommand{\shorttitle}[1]%
{\markboth{Proceedings of the 31\MakeLowercase{$^{st}$} ICRC, {\L}\'{o}d\'{z} 2009}{#1} }
\newcommand{\etal}{\MakeLowercase{\textit{et al. }}} 

\begin{document}
\title{Lunar gamma-ray emission observed by {\it Fermi}}

\author{\IEEEauthorblockN{Nicola Giglietto\IEEEauthorrefmark{1} on behalf of the {\it Fermi}-LAT Collaboration
			  }
                            \\
\IEEEauthorblockA{\IEEEauthorrefmark{1}Dipartimento Interateneo di Fisica di Bari and INFN Bari
     }
}

\shorttitle{N.Giglietto \etal {\it Fermi} Moon observation}
\maketitle

\begin{abstract}
 \emph{Fermi} LAT is performing an all-sky gamma-ray survey from 30 MeV to 300 GeV with unprecendented senstivity and angular resolution.
\emph{Fermi} has detected high-energy gamma rays from the Moon produced by interactions of cosmic rays with the lunar surface.  This radiatioon was previously observed by EGRET on CGRO with significantly lower statistical significance. We present the lunar analysis for the first six months of the Mission and showing images of the lunar gamma-ray emission.  We also compare the flux measurements  with models the earlier EGRET measurements.
  \end{abstract}

\begin{IEEEkeywords}
 Gamma-ray astronomy, Moon, cosmic-rays
\end{IEEEkeywords}
 \begin{figure}[hbt]
   \centering
  \includegraphics[height=3.1in,width=3.in]{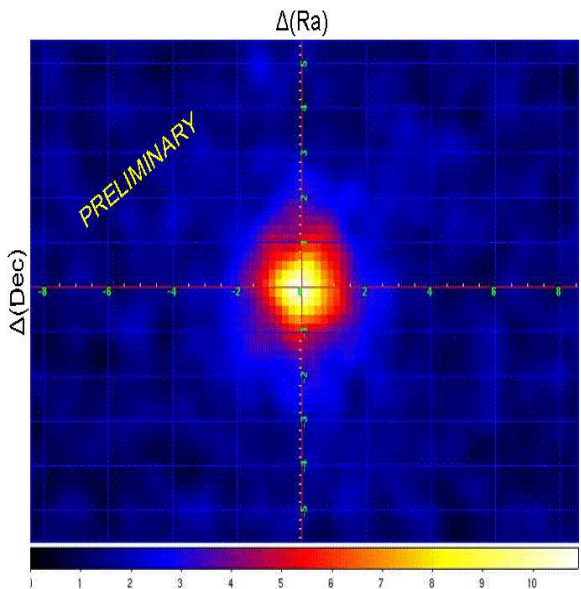}
  \caption{Count map in a Moon centered frame, Right Ascension and Declination offsets in degrees respect to the Moon position in abscissa and ordinate respectively.
 The offset ranges are $\pm~8^\circ$ both for Right Ascension and $\pm~5^\circ$ for Declination.
  The image has been obtained by using photons with E$>$~100~MeV, a bin width of 0.2 degrees with a gaussian smoothing 2 bin radius.  
    The z  colour scale is linear with the counts.}
  \label{moon-map}
 
\end{figure}

\section{Introduction}
 \emph{Fermi} was launched from Cape Canaveral on the 11th of June 2008.
 It is currently in an almost  circular orbit around the Earth at
 an altitude of 565~km having an inclination of 25.6$^\circ$ and an orbital period of 96 minutes. After an initial activation period and on-orbit calibration \cite{LATcalib}, the observatory was  put into a sky-survey mode.
The observatory has two instruments onboard,
the Large Area Telescope\cite{LATpap} (LAT),  a pair-conversion gamma-ray detector with a tracker   
and calorimeter, and a Gamma Ray Burst Monitor (GBM) that is dedicated to the detection of gamma-ray bursts. The two instruments  on \emph{Fermi} provide coverage over  the energy range from a few kev to several hundreds GeV. 

\begin{figure*}[tbh]
  \centering 
\includegraphics[height=3.75in,width=5in]{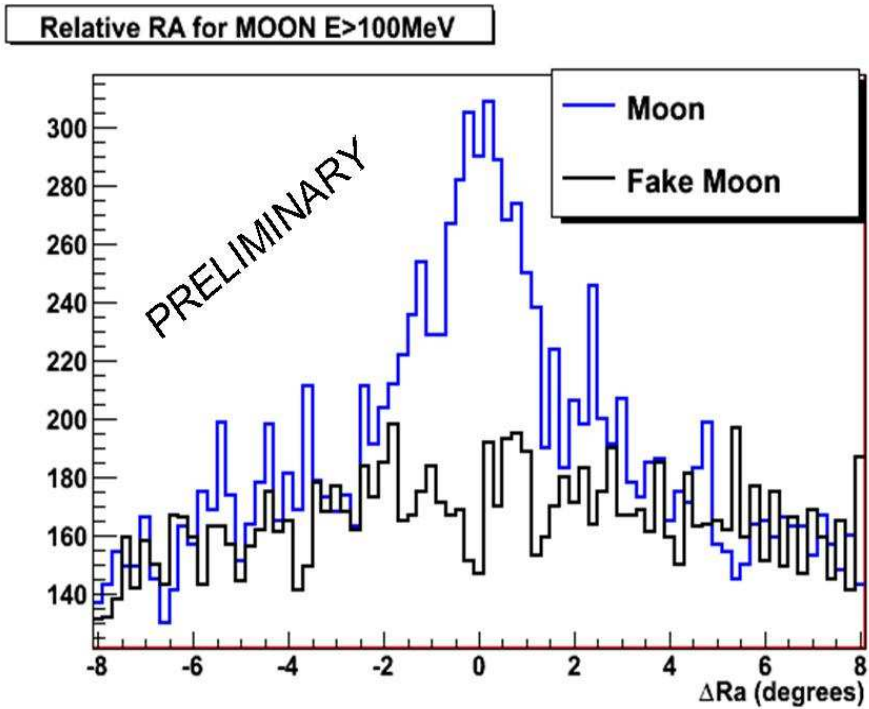} 
   \caption{Count map of the events with an angular distance of 15$^\circ$ from the Moon and with E$>$~100~MeV, as a function of Right Ascension offsets in degrees respect to the Moon position. Superimposed as black line the fake Moon count map distribution.
 Observed data are consistent with the expected angular resolution for 100~MeV photons\cite{irfs}.}\label{Moon-ra}
  \end{figure*}
\begin{figure*}[bth]
  \centering 

  \includegraphics[height=3.75in,width=5in]{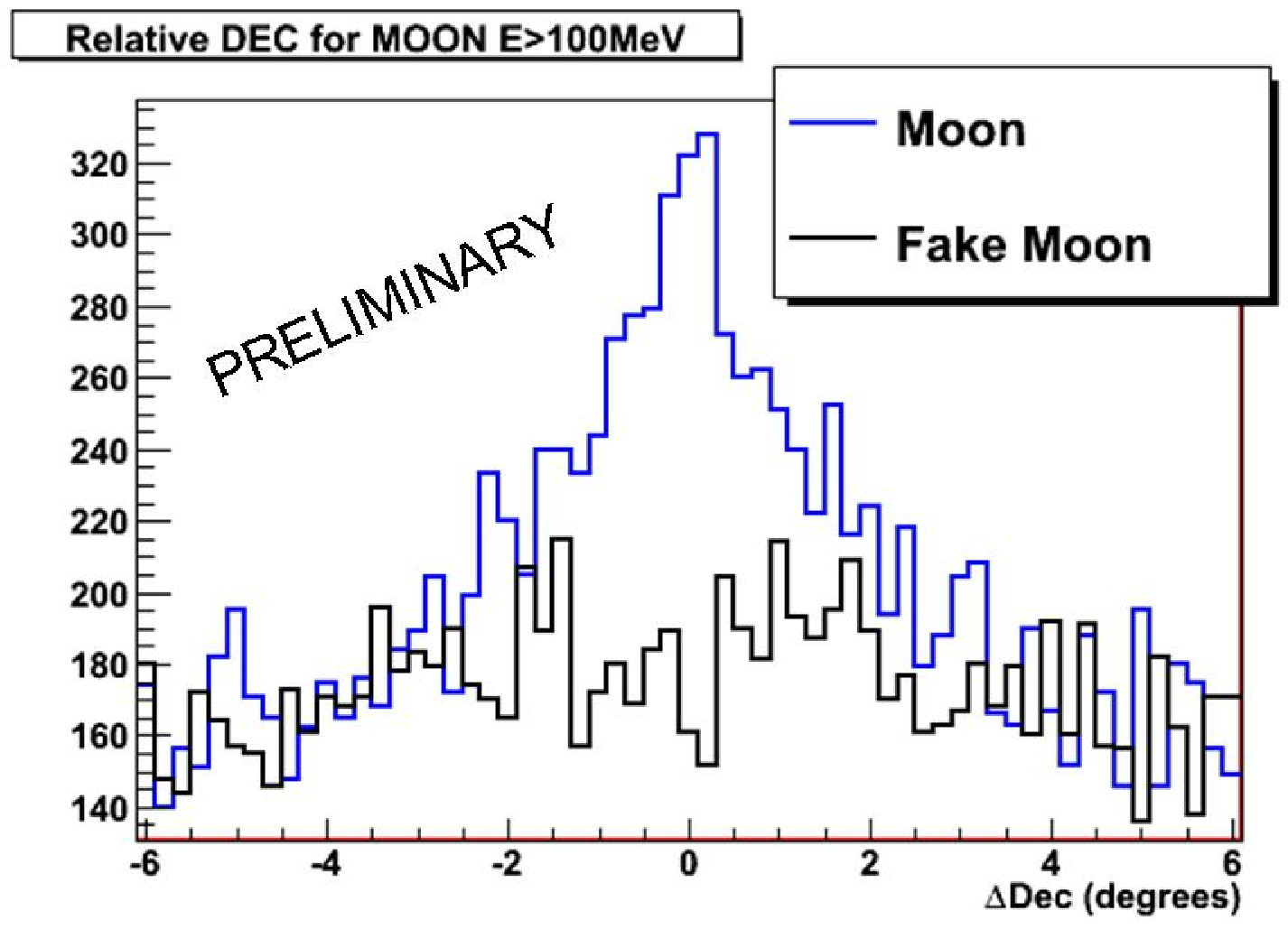} 
             
   \caption{Count map of the events with an angular distance of 15$^\circ$ from the Moon and with E$>$~100~MeV, as a function of Declination offsets in degrees respect  
to the Moon position. Superimposed as black line the fake Moon count map distribution.
Observed data are consistent with the expected angular resolution for 100~MeV photons\cite{irfs}.}
\label{Moon-dec}
 \end{figure*}

Here we report the detection and measurements of gamma-ray emission from the Moon resulting from cosmic-ray interactions with Moon.  The $\gamma$-rays results from decays of neutral pions and kaons.  Cosmic-ray interaction with the lunar surface is well established and the gamma-ray spectrum has been computed \cite{Morris84}; more recent calculations have been performed using a detailed description of the regolite lunar surface \cite{Igor} and  GEANT4\cite{GEANT4}
taking into account all the interactions with the specific composition of the lunar rock (\cite{Igor, Igor2}).
These calculations indicate that
the kinematics of the collisions of cosmic rays  hitting  the lunar surface, produce a secondary 
particle cascade that develops deep in the rock. A small fraction of the secondary pions that are low energy are directed toward to lunar surface and decay to produce a soft $\gamma$-ray spectrum.
 Moreover the same work  indicates that
 the lunar gamma-ray spectrum should exhibit a narrow 67.5 MeV line due to neutral pions decaying at rest and a steep spectrum. Finally
since the high energy $\gamma$-rays can be produced by cosmic rays hitting the lunar surface with almost tangential traiectories, the limb of the disk surface should give the larger contribution to the emission. 

The discovery of $\gamma$-ray emission from the Moon was made during early analysis of EGRET data \cite{EGRET} and provided a measurement of the gamma-ray spectra from cosmic-ray
 interactions with the lunar surface. They found an integral flux F(E$>$100~MeV)=$(4.7\pm0.7)\times 10^{-7}~cm^{-2}~s^{-1}$; this was confirmed in recent work\cite{Orlando08}. 

With its large effective area the \emph{Fermi}-LAT should be able to explore in detail the features of the lunar emission.  In this paper we report first results from analysis of the first seven months of observations, updating previous Fermi observations of lunar gamma-ray emission\cite{Giglietto,Brigida}; we also present preliminary measurements of observed fluxes.

\section{Data selection}

The data used in this study were collected from 4 August 2008  to the end of February 2009 (7months) when the Sun was at its minimum of activity. We have
applied a zenith cut of 105$^{\circ}$ to eliminate photons from the Earth's limb.  We used the "Diffuse" class
\cite{LATpap}, corresponding to events with the highest probability of being true photons, for this analysis. We also used Science Tools version v9r11 and IRFs (Instrumental Response Functions) version P6\_V3. The position of the Moon has been  computed using a JPL\cite{jpl} library interface and parallax corrections are taken into account. 
  Photon events for analysis were selected in a moving frame of 15 degree radius and centered on the istantaneous Moon position. We used the unbinned  likelihood analysis technique typically used for astrophysical sources but, because the Moon move quickly across the sky, we had to take additional precautions to determine the emission spectrum. 
In order to avoid strong variations of background photons, we excluded time intervals when the Moon was close to the galactic plane or bright sources. 
We therefore required that the  Moon was at least  30$^\circ$ from the galactic plane i.e. $|B_{moon}|>30^\circ$, and we remove also any time intervals in which any individual bright object has an angular distance less than  5$^\circ$ from the Moon. As an additional test of our background rejection we repeated the analysis using a 'fake' moon that followed the lunar trajectory but was shifted 30 degrees behind the true trajectory.    The 'fake' Moon was therefore exposed to the same
celestial sources as the true Moon.



\section{Results}

Fig.~\ref{moon-map} shows the count map of photon events with E$>$~100~MeV in offsets of celestial coordinates relative to the Lunar position. Emission from the Moon is clearly visible and centered on the expected location in this relative coordinate frame.

Fig.~\ref{Moon-ra} shows the count map of the events with E$>$~100~MeV and within a 15$^\circ$ from the Moon center projected onto Right Ascension while in Fig.\ref{Moon-dec} the events are projected onto Declination. The coordinates are  offsets in degrees of celestial coordinates relative to the Moon position. In these figures the counts observed for the 'fake' Moon are superimposed and demonstrate the need to carefully consider
the  background before any lunar analysis can be performed. Simulated data confirm that the observed shape is in agreement with a pointlike source and the calculated point spread funtions of the LAT\cite{LATpap,irfs} for the expected lunar gamma-ray spectrum.
In a more detailed 
analysis of a  one-year database we will perform more detailed analysis of the shape of emission  to determine its true extent and whether we can discern evidence for the expected limb brightening.

 

To study the spectrum of the gamma-ray emission we have used the standard unbinned maximum-likelihood spectral estimator provided with the LAT science tools {\tt gtlike}. This preliminary analysis  was performed by fitting the ''fake'' moon data to model the background and the Moon data sample
with either a simple power law or other functional forms like a log-parabola summed to the background model. 

The fitted values obtained using a simple power law give a good test-statistics value of 4320.3 (defined as $TS=2(\log L-\log L_0)$, being $L$ and $L_0$ the likelihood values when the source is considered or not)     and indicate a power law index of $-3.13\pm0.03$.
As a result of these fits we estimate the observed flux as $F(E>100~MeV)=1.1\pm0.2\times 10^{-6}$ $cm^{-2}s^{-1}$.
The error includes
the estimation of overall systematic error of about 20$\%$ for these measurements, due essentially to the uncertainties of the detector effective area and due to  
the inefficiencies in gamma-ray detection due to pile-up effects from near coincidences with cosmic rays in the LAT detector.

However since the expected spectrum\cite{Igor,Igor2} should exhibit more complex features we plan to  more extensively study the spectral shape when the statistical sample is larger and the instrument response below 100 MeV is better determined. 


\section{Conclusions}
The gamma-ray emission from the Moon discovered by EGRET  has been confirmed by \emph{Fermi} and agrees in intensity for emission models that take into account the level of solar modulation. Our preliminary flux estimation for the lunar $\gamma$-ray emission is $F(E>100MeV)=1.1\pm0.2\times 10^{-6}\,cm^{-2}s^{-1}$ with a spectral index of $-3.13\pm0.03$ obtained by fitting a simple power law between 100~MeV to 1~GeV. 
We expect that  with a larger sample of events available after accumulating data for one year,  \emph{Fermi} should be able to explore the other features of the lunar spectrum, e.g. the $\pi_0$ peak and also spatial structure on the lunar disk.

\section*{Acknowledgements}
The \emph{Fermi} LAT Collaboration acknowledges support from a number of agencies and institutes for both development and the operation of the LAT as well as scientific data analysis. These include NASA and DOE in the United States, CEA/Irfu and IN2P3/CNRS in France, ASI and INFN in Italy, MEXT, KEK, and JAXA in Japan, and the K.~A.~Wallenberg Foundation, the Swedish Research Council and the National Space Board in Sweden. Additional support from INAF in Italy for science analysis during the operations phase is also gratefully acknowledged.

\vspace{1,5cm}

\end{document}